# First-principles study of the Mn, Al and C distribution and their effect on the stacking fault energies in austenite

N. I. Medvedeva, [1,2], M. S. Park [1], D. C. Van Aken, [1] and J. E. Medvedeva [1]

[1] Missouri University of Science and Technology, Rolla, MO 65409

[2] Institute of Solid State Chemistry, Yekaterinburg, Russia

E-mail: medvedeva@ihim.uran.ru

**Abstract**

We present *ab-initio* simulation of manganese, aluminum and carbon impurities in austenite and demonstrate their inhomogeneous distribution, which involves the repulsion of interstitial carbon atoms, the formation of bonded Mn-C pairs as well as a short range Al-ordering of $D0_3$-type. The mechanisms for the formation of stacking faults in Fe-Mn-Al-C are considered, and we find that the impurities have influence on the stacking fault energies only when located within a few interatomic layers near stacking fault. As a result, the stacking fault energy does not depend on the average concentration of impurities in matrix, but is highly sensitive to the concentration of the impurities in the vicinity of stacking fault defect. We predict that manganese shows a slight tendency for segregation near SF, while carbon prefers to be located far from the stacking fault region. Both aluminum and carbon impurities linearly increase the SFE, while the formation of Mn-C pairs and short range Al-ordering restrain the SFE growth. Short range order in Fe-Al-C alloys strongly affects the energy barrier for nucleation of dislocations and may lead to softening phenomenon.

# 1. Introduction

The deformation behavior of high manganese steels such as Hadfield steel has been a subject of intensive investigations for many years [1-6]. The high work hardening rate in these austenitic steels was related to dynamic strain aging and the formation of Mn-C dipoles in the face centered cubic (fcc) solid solution [1] or to the deformation twins formed during plastic straining [2]. More recent studies [3,4] showed that the deformation-induced twin boundaries are as effective in work hardening as the dislocation accumulation. The active deformation mechanism associated with enhanced plasticity and high strength (martensite formation, twinning or glide dislocations) in the high manganese austenitic alloys is controlled by the intrinsic stacking fault (ISF) energy [12,13]. This stacking fault can be produced by the nucleation and glide of Shockley partials having a/6<112> Burgers vectors that create a hexagonal close packed region, which is equivalent to the crystal structure of ε-martensite. The ISF energy (ISFE) plays a central role in forming ε-martensite and deformation twins, and observation of either transformation induced plasticity (TRIP) or twinning induced plasticity (TWIP) depends on ISFE [18,19]. Deformation induced ε-martensite occurs for ISFE below 18 mJ/m$^2$, whereas deformation twinning may occur between 12 and 35 mJ/m$^2$ [14]. Twinning is delayed to a higher critical shear stress, or a greater strain, as the ISFE is increased [10,11] and planar slip forming microbands dominates at higher ISFE (90 mJ/m$^2$) as evident in fully austenitic Fe-28Mn-10Al-1C steel [20].

Substitutional and interstitial impurities strongly affect the deformation mechanism by changing the SFE. The effect of solute on SFE is rather complex and depends on solute concentration, temperature, grain size, magnetism and chemical interactions, which play an important role. Intrinsic stacking fault energy, $\gamma_{ISF}$, is usually estimated from the thermodynamic model as $\gamma_{ISF} = 2\rho\Delta G^{\gamma-\varepsilon} + 2\sigma^{\gamma/\varepsilon}$, where $\Delta G^{\gamma-\varepsilon}$ is a difference in the Gibbs free energies of γ-austenite and ε phases, which are estimated from the regular solution model for multicomponent systems; ρ is the atomic density on the (111) plane, and $\sigma^{\gamma/\varepsilon}$ is the interfacial energy between γ and ε phases [16, 21]. Such estimations have demonstrated that carbon and aluminum strongly increases the SFE [8, 21], while the effect of manganese is more complex. The linear increase of SFE with manganese concentration up to 50 wt.% was obtained in [22,23], while the parabolic dependence with decreasing of SFE at the low Mn concentration and increasing at the high Mn concentration was predicted in [24-26] where the SFE minimum corresponds to 12 at.% [24], 15 at.% [25] and 22 at.% Mn [26].

The above simple solution model, which is used to predict impurity effect on $\Delta G^{\gamma-\varepsilon}$, does not adequately take into account the changes in the structure (relaxation) and in the interatomic interactions caused by alloying and, therefore, does not provide the atomistic origin of the SFE dependence on the impurity. For this, first-principles methods, which have been successfully used to study the structural and magnetic phase stability of Fe [27-29], Fe-Mn [30,31], Fe-C [32-35], and Fe-Mn-C [35,36], were recently employed to calculate the SFE in γ-Fe [37-38], Fe-N [39], Fe-Mn [38-40] and Fe-Cr-Ni, Fe-Cr-Ni-Mn, Fe-Cr-Ni-Nb alloys [41,42]. Although first-principles approaches give the results at zero temperature, they provide important atomistic information on the changes in the SFE with alloying. The ab-initio calculations [37-40] predict the SFE in fcc Fe to be negative, in agreement with thermodynamic instability of austenite against ε-martensite. Nitrogen was found to increase the SFE by 73 mJ/m$^2$ per 1 at.% N, while manganese decreases the SFE at a rate of 3 mJ/m$^2$ per 1 at.% Mn up to concentration of 8 at.% Mn [39]. However, modeling of the manganese effect in austenitic $Fe_{1-x}Mn_x$ alloys using ordered structures of Fe, $Fe_{75}Mn_{25}$, $Fe_{50}Mn_{50}$, $Fe_{25}Mn_{75}$, and Mn, predicts a monotonic increase in the SFE and the thermodynamic instability of austenite phase up to 100% Mn [38]. These results contradict experimental and theoretical findings on the austenite to ε –martensite transformation in $Fe_{1-x}Mn_x$. Recent calculations for chemically disordered Fe-Mn alloy [40] demonstrated that the non-linear dependence of SFE with the SFE minimum at 20 at.% Mn and 40 at.% Mn may be reproduced for nonmagnetic and paramagnetic states, respectively. However, these calculations (for ordered and nonmagnetic, for disordered and nonmagnetic, as well as for disordered and paramagnetic alloys) predict the SFE to be lower than -150 mJ/m$^2$ for all Mn concentrations. As was noted earlier [40], this discrepancy may appear due to thermal effects, crystal defects, local deformations as well as due to interstitial impurities (e.g. C and N), which are always present in steel.

It should be noted that interstitial carbon strongly increases the SFE and suppresses twinning, while the role of manganese with respect to intrinsic stacking fault energy (ISFE) is more complex: it is largely responsible for a wide variation in microstructure, work hardening behavior, and enhanced plasticity, e.g. Refs. [14-17]. These impurities may form bound Mn-C pairs in Fe-Mn-C alloy that may significantly retard the dislocation motion [28,29]. Aluminum is an effective alloying element to raise the intrinsic stacking fault energy (ISFE) of high Mn steels and to suppress formation of deformation twins, but these alloys still maintain high work hardening rates [6-9] which was associated to planar slip and the formation of high dislocation density sheets [8]. There is strong evidence [5] that short range order (SRO) is important in the work hardening behavior prior to the formation of twins in Hadfield steel. Effects of short range order and the formation of Mn-C dipoles on ISFE in Fe-Mn-Al-C alloys have not been studied using either thermodynamic models or ab-initio methods. Since the interaction between solute and interstitial atoms plays an important role in the mechanical properties, the knowledge of the impurity distribution is critical for understanding the microscopic origin of the impurity effect.

In this paper, we use an *ab-initio* approach to study the distribution of carbon with respect to the substitutional manganese and aluminum, as well as the location of these impurities with respect to each other in austenite. The local crystal distortions and variation in lattice parameters were accurately determined via the atomic force and total energy minimization. Based on these results, we calculated the generalized stacking fault energies (GSFE) which represent the energies for sliding of atomic planes. We study the influence of manganese, carbon and aluminum on the stacking fault energies by considering their different positions and concentrations at the stacking faults.

We employed the Vienna ab-initio simulation package (VASP) with the projector augmented waves (PAW) for pseudopotentials [43,44]. The generalized gradient approximation (GGA) of Perdew et al. [45] was used for the exchange-correlation functional. Energy cutoff of 350 eV was used for the plane-wave basis set. Simulation of impurity distribution in fcc Fe was performed for supercells of 32 atoms, while for GSFE calculations we used a 24-atom supercell. For the calculations involving both the 32- and 24-atoms supercells, a 4x4x4 $k$-point mesh was chosen. The atomic positions were relaxed until the forces were smaller than 0.01 eV/Å. Despite the importance of the magnetic interactions in Fe-based alloys, we are restricted by the calculations for nonmagnetic state in this paper and discuss only the role of interatomic interactions.

## 2. Distribution of interstitial carbon and substitutional Mn and Al impurities in austenitic Fe-Mn-Al-C alloys

For nonmagnetic (NM) fcc Fe, we obtained the optimized lattice parameter of 3.456 Å which is in accord with previous ab-initio calculations [27,35]. Interstitial octa-site carbon at 3 at.% concentration slightly increases the lattice parameter, up to 3.472 Å. We compared the total energies for supercells with two carbon atoms in different octahedral sites (Fig.1a, sites 1-5) and found that they prefer to occupy the most distant sites from each other (Table 1). The configuration with two nearest carbon atoms forming the 180$^o$ C-Fe-C pairs is the least likely. The density functional theory [46] and molecular dynamic [47] calculations gave similar results for carbon distribution in bcc Fe-C, where the C-Fe-C atomic configuration was predicted as the least probable. Previous estimations of the C-C interaction energy in the austenitic Fe-C solid solutions within statistical methods also demonstrated that carbon atoms repel each other and the strength of repulsive interaction decreases with increasing temperature [48]. The C-C repulsive interaction strongly affects the carbon diffusion making it dependent on the carbon concentration [49,50]. It should be noted that the C-C bonds are also weak in metastable cementite $Fe_3C$ [33,49], which is formed in carbon steel and plays an important role in strengthening. Because of the weak C-C and Fe-C bonds, carbon non-stoichiometry and carbon occupation of octahedral interstices against the normal trigonal sites have little effect on the electronic structure and the formation energy of cementite [51].

For Fe-Mn-C, for which we modeled using a Fe31MnC supercell, we found that the formation of a Mn-C pair has the lowest energy (Fig.1b) among the Mn and C positions (2, 3 and 4). The second manganese atom (Fe30Mn2C supercell) prefers to occupy the position 1 and to form the 180° Mn-C-Mn chain (Fig.1c). The next possible position is site 2, while the other positions for manganese (3-11) have higher energies. The Mn-C binding energy, estimated as the energy difference between the nearest and distant carbon positions relative to manganese, increases with Mn content from 29 meV (3 at.%Mn) to 85 meV (6 at.%Mn), see Table 1. Thus, the Mn-C interaction is attractive and may slow the carbon diffusion during carburization of high manganese steel. The difference between configurations with free carbon distribution and with carbon trapped by manganese is rather small for the low Mn and C concentrations and both carbon sites are possible in austenitic Fe-Mn-C alloy. We would like to stress that here we discuss the nonmagnetic results, and the local magnetic interactions may increase the Mn-C-Fe binding energy, as shown for ferrite Fe-Mn-C alloys [36].

Our calculations for Fe-Al-C (Fe30MnAlC supercell) predict that the most energetically favorable substitutional position for aluminum is site 2 (second coordination shell with respect to carbon, Fig.2). Strong preference of aluminum occupation of the site 2 points out the short-range order in the Fe-Al-C alloy. The predicted local distribution of Al and C in austenite corresponds to the E21 structure of κ-carbide $Fe_3AlC$, which is precipitated in the Fe-Al-C alloys at aluminum and carbon concentrations higher than 5% and 0.3 %, respectively, and plays a crucial role in the mechanical properties of these alloys. Homogenous precipitation and dispersion of nano-size κ-carbides are the main mechanisms of hardening and enhanced ductility of Fe-Mn-Al-C alloys [52,53]. Evidence of Fe-Al-C clustering is experimentally supported by Park et al. [10], where they show electron diffraction patterns with perovskite superlattice reflections or the κ-carbide ($Fe_3AlC$). We also examined the position of Al with respect to Mn-C pair (Fig. 2b) and found that aluminum prefers to be in site 3, while substitution at site 9 next to the Mn atom is higher by 120 meV. This result shows that Al atoms avoid manganese as a nearest neighbor.

The changes in crystal structure induced by impurities were investigated using the force and total energy minimization. Carbon inserted in the octahedral interstitial site increases the lattice parameter from 3.456 Å to 3.472 Å for 3 at.% concentration (Fe32C), that is in a good agreement with experimental measurement of the fcc lattice parameter with varying carbon content [54]. The nearest Fe atoms move outward from carbon and the Fe-C distance increases from 1.73 Å to 1.87 Å under the atomic relaxation. Thus, we obtained that carbon in fcc Fe-C uniformly expands the lattice and the distances between octahedral-site carbon and the nearest neighbor iron atoms increase by 7.7% as compared with those in pure fcc Fe. Manganese substituted for Fe next to octahedral-carbon distorts the symmetry of Fe5MnC octahedron. The Fe-C distances for planar iron remain the same as in Fe-C, while the Mn-C and Fe-C distances increases (to 1.90 Å) and decreases (to 1.87 Å) in the 180º Mn-C-Fe pair, respectively. There is a stronger bonding between carbon and two Mn atoms in the octahedron Fe4Mn2C, where all Fe(Mn)-C distances are equal and the octahedron is undistorted.

Thus, we find that manganese, carbon and aluminum impurities have energetically preferable relative positions, and the formation of bonded Mn-C pairs and short ordered structures such as κ-carbide is likely in austenitic Fe-Mn-Al-C alloys. As a result, the carbon distribution where nearest carbon atoms and octahedral interstitials close to aluminum are avoided, as well as the Mn-C binding is expected to suppress the carbon diffusion. As mentioned in the Introduction, short range order in the Fe-Mn-C steels alloyed with aluminum has been suggested to promote planar slip as a result of a glide plane softening phenomenon, whereas the formation of Mn-C dipoles is a reason for the high work hardening rate.

In the next section, we consider how these impurities and their distribution affect the stacking fault energies and discuss possible changes in the deformation behavior.

## 3. Effect of Mn, C and Al on the generalized stacking fault energies

In fcc metals, a slip <112>(111) and a planar shift at the Burgers vector $\mathbf{b}_p=1/6<112>$ produces a minimum on the GSF curve (Fig. 3) which corresponds to the stable intrinsic stacking fault (ISF). Intrinsic faults are formed by planar shear on the (111) plane which changes the atomic packing from ABCABCAB to ABCACABC, and the associated energy, $\gamma_{ISF}$, suggests the formation of hcp layers. Alloy additions that decrease $\gamma_{ISF}$ would stabilize the hcp phase and favor the $\gamma \rightarrow \varepsilon$ transformation, while those that increase $\gamma_{ISF}$, suppress the formation of ε-martensite. The displacement at $1/12<112>$ ($0.5|\mathbf{b}_p|$) corresponds to unstable stacking fault (USF) and this energy $\gamma_{US}$ controls the barrier for the formation of stacking faults as well as the dislocation nucleation. The third important energy characteristic of the planar shear is the displacement at $1/3<112>$ ($2|\mathbf{b}_p|$) and corresponds to the maximum in the GSFE curve ($\gamma_{MAX}$), which may signify a barrier for the dislocation movement. These GSF energies are the key parameters that determine the structure and mobility of dislocations [55].

We modeled the stacking fault defects using a 24-atom supercell (six layers with four atoms per layer) and calculated the GSF energies as the total energy change caused by a rigid shift of a half of the crystal along the <112> direction in the (111) slip plane. The calculated GSF energies for fcc Fe (Fig. 3), $\gamma_{US}$ = 439 mJ/m$^2$ , $\gamma_{ISF}$ = -347 mJ/m$^2$, $\gamma_{MAX}$ = 1932 mJ/m$^2$ are in agreement with previous *ab-initio* results [37-39]. The negative value of $\gamma_{ISF}$ indicates a preference of hcp structure over fcc for pure Fe at low temperature and it means that austenite is thermodynamically unstable relative to ε-martensite.

To study the dependence of $\gamma_{ISF}$ on manganese concentration, we performed calculations for a different number of Mn atoms at the stacking fault on the (111) plane. Our calculations provide parabolic dependence of $\gamma_{ISF}$ with manganese content at the SF (Fig. 4). The SFE decreases with a mean rate of 5 mJ/m$^2$ per 1 at.% within the interval from 0 to 12 at.% Mn, and increases with a slightly larger rate of 6.6 mJ/m$^2$ per 1 at.% from 15 to 33 at.% Mn. The minimum corresponds to 13 at.% Mn and coincides with the predictions made in Ref. [24,25].

As it follows from Fig. 4, the Mn concentration range up to 30% yields smaller values of the SFE as compared to those in fcc Fe. A decrease in SFE favors the formation of martensite. Thus, our calculations suggest that the martensite structure may be formed within this concentration interval, whereas austenite is stable for higher Mn concentrations. These results correlate with the phase diagram of Fe-Mn which demonstrates that ε–martensite is not formed above 30% Mn. Extrapolation of the SFE curve by the function y=0.2927*x$^2$-8.09*x-346 (Fig. 4) gives that $\gamma_{ISF}$ changes sign from negative to positive at ~50 at.% Mn.

The parabolic behavior of $\gamma_{ISF}$ was obtained within the nonmagnetic scheme, and we can suggest that the interatomic interactions but not magnetism determines the nonlinear dependence of the SFE on Mn content. Note, that we considered variations in manganese concentration exactly at the stacking fault. Earlier first-principles calculations, which predicted a gradual increase in the SFE for all Mn concentrations [38], were performed for nonmagnetic Fe (A1), Fe$_{75}$Mn$_{25}$ (L1$_3$), Fe$_{50}$Mn$_{50}$ (L1$_0$), Fe$_{25}$Mn$_{75}$ (L1$_3$), and Mn (A1) phases where manganese atoms were in the ordered positions. In order to determine how manganese positions relative to the SF region influences $\gamma_{ISF}$, we also calculated the following two configurations: (i) when the *n* manganese atoms were distributed not on the SF plane, but through the *n* layers below the stacking fault; and (ii) when one manganese atom was substituted in the $n^{st}$-layer (*n* = 0, 1, 2, 3) below SF. Such calculations will provide an answer to whether $\gamma_{ISF}$ depends on the Mn position relative SF or on the total Mn concentration in bulk.

In the first case (Fig. 4), we found that the SFE does not depend on Mn concentration for x > 10 at.%, and all changes in SFE are produced by the manganese atoms located in the 0$^{th}$- and 1$^{st}$-layers below stacking fault. Indeed, as follows from the calculations (ii), manganese in the second layer below SF does not affect $\gamma_{ISF}$ (we obtained -372, -354 and -349 mJ/m$^2$ for manganese in the 0$^{th}$, 1$^{st}$ and 2$^{d}$ layer, respectively). Hence, manganese that is distant from the stacking fault layer does not change the SFE ($\gamma_{ISF}$ = -347 mJ/m$^2$ for fcc Fe). Thus, it is the manganese atoms distributed within the one interlayer distance near the SF region that affect $\gamma_{ISF}$, but not the total manganese concentration in the Fe-Mn alloy.

The lowering of SFE with increasing $n$ also suggests that the location of manganese near the stacking faults is energetically more preferable than being far away from this region. To elucidate the effect of manganese on the formation of stacking faults, we calculated the unstable stacking fault energies USFE ($\gamma_{US}$), which represent the lattice resistance to the formation of stacking faults. As shown in Fig. 4, $\gamma_{US}$ decreases with increasing Mn content at SF for x>20 at.% Mn. This points to a lower energy barrier for the formation of stacking faults in the regions with high manganese concentration. These results indicate the possibility of manganese segregation near stacking faults and, hence, near the boundary of the fcc-hcp phase separation.

We would like to stress that the magnitude of SFE obtained in our study, as well as in previous *ab-initio* calculations, is greatly differ from the experimental values and the thermodynamic estimations. For example, SFE for pure Fe being extrapolated to T=0 K is higher than -40 mJ/m$^2$ [40]. This discrepancy may be related to magnetic interactions. Indeed, a strong temperature dependence of the SFE was related to magnetism. [56]. The comparison of paramagnetic and nonmagnetic results for pure Fe showed that magnetism increases the SFE from -350 mJ/m$^2$ to -150 mJ/m$^2$ [40], but nevertheless it is far from the extrapolated value. The effect of light interstitial impurities is a possible reason for this discrepancy. In particular, *ab-initio* calculations predict that nitrogen increases SFE by 73 mJ/m$^2$ per 1 at.% N and that SFE is close to zero at 4 at.% N [39].

Next, we consider the effect of carbon occupying the octahedral interstitial sites in the $n^{st}$-layer ($n$ = 0, 1, 2, 3) below SF in fcc Fe. One carbon atom at the SF layer ($n$ = 0) increases both $\gamma_{US}$, and $\gamma_{ISF}$, but they are gradually reduced to the values for pure Fe as $n$ increases (Table 2). Carbon in the third layer below SF ($n$ = 3) has a little influence on the stacking fault energy. There is a strong dependence of $\gamma_{ISF}$ on both the carbon location and its concentration, whereas the variations in $\gamma_{US}$ are less significant. A much larger increase would occur in the case when two nearest carbon atoms are at the SF layer ($n$ = 0,0). However, as demonstrated above, the C-C interaction is repulsive and carbon atoms tend to be located far apart in austenite. Therefore, the probability of the configuration $n$ = 0,0 is small. The increase in $\gamma_{US}$ (energy barrier for the SF formation) with carbon content (Table 2) also indicates that the segregation of interstitial carbon at the stacking fault is unlikely.

The dependence of the stacking fault energy on the impurity layer $n$ shows (Table 2) that carbon, in marked contrast to manganese, has a strong effect within a more extended region, which covers three atomic layers near the stacking fault defect. We averaged $\gamma_{ISF}$ on the three SF planes ($n$ = 0, 1, 2) and obtained the increase of 74 mJ/m$^2$ per 1 at.% C. This coincides with the nitrogen effect on SFE that was predicted to be 75 mJ/m$^2$ per 1 at.%N [39]. These results imply that in the Fe-C alloy, the fcc phase is more favorable than hcp. Indeed, carbon suppresses the $\gamma \rightarrow \varepsilon$ transformation and is known as a strong austenite stabilizer. The decrease in SFE obtained from the thermodynamic estimations (~12 mJ/m$^2$ per 1 at.% C) is notably smaller than our value obtained via averaging through three n layers near the SF plane. The high $\gamma_{US}$ shows that carbon does not segregate at SF and does not favor the formation of stacking faults. Carbon prefers to be far from the stacking fault region, and this may be the reason for the overestimated value obtained by including the nearest region to SF into the average. Taking into account this argument, we averaged $\gamma_{ISF}$ through the more distant layers ($n$ = 2, 3) and obtained the increase of 28 mJ/m$^2$ per 1 at.% C, which is in a better agreement with thermodynamic estimations.

In accord with our findings on the favorable impurity positions, we study how the formation of Mn-C pair may influence the SFE. We performed calculations for one (Mn-C-Fe) and two manganese (Mn-C-Mn) atoms substituted in the preferable positions near the interstitial carbon at stacking fault and obtained that the ISFE decreases by 5 mJ/m$^2$ and 9 mJ/m$^2$ per 1 at.% Mn in comparison with the value for Fe-C, respectively. This means that the formation of bound Mn-C pairs near the stacking fault plane restrains the increase in SFE caused by carbon alone. As a result, the calculated SFE is in a better agreement with the observed value.

For aluminum, we considered the substitutional sites in the $n^{st}$-layer ($n$ =0, 1, 2) below SF as well as the Al concentration effect (Table 2, Fig.5). We obtained that Al impurity in fcc Fe lead to an increase of SFE which depends on the aluminum distribution and concentration. Aluminum (4 at.%) located at SF or in the first (nearest) layer leads to the similar increase of $\gamma_{ISF}$, while there is no effect on SFE when Al is in the second or third layer ($n$ =2 or 3). The increase in $\gamma_{ISF}$ averaged over the layers $n$ =0, 1, and 2 is 10 mJ/m$^2$ per 1 at.% Al. The concentration of 8 at.% Al was modeled by two atoms which were: (i) both at SF ($n$ =0,0), or (ii) one atom at SF and one atom in the first layer ($n$ =0,1) below SF. For Al atoms in $n$ =0,0 and $n$ =0,1 positions, the SFE increases by 19 mJ/m$^2$ and 11 mJ/m$^2$ per 1 at.% Al, respectively. This result shows a correct trend in the changes of $\gamma_{ISF}$ with Al addition, but shows a much greater increase than previous thermodynamic estimations and experimental data, that showed the increase in SFE of 5 mJ/m$^2$ [21].

We found that $\gamma_{US}$ is sharply reduced with aluminum alloying (Fig. 5). This means that aluminum may segregate near the stacking fault plane. Secondly, it points out that aluminum sharply lowers the energy barrier for stacking faults and dislocation formation, and therefore favors enhanced plasticity. As mentioned in the Introduction, a planar glide before mechanical twinning was observed in the Fe–Mn–Al–C alloys with rather high SFE, while a planar glide is traditionally considered to dominate in alloys with low SFE [10]. Here it should be noted that planar slip in face centered cubic materials is often associated with short range order or short range clustering. Gerold and Karnthaler [11] explained the softening as a result of disorder of either short range order or short range clustering by the initial dislocation glide during yielding, which permits easy flow of successive dislocations. A short range ordering suggested in Fe–Mn–Al–C steel was related to the precursor of $Fe_3AlC_x$ precipitates [10].

Our investigations of the relative distribution of impurities in fcc Fe demonstrate the preference for aluminum to occupy the position related to $D0_3$ structure. In the Fe-C alloy it corresponds to the formation of perovskite-like κ-carbide $Fe_3AlC$ with $E2_1$ crystal structure, where aluminum occupies the (0,0,0) position, iron atoms are in the (½ , ½, 0) positions, and carbon is located at the center octahedral site (½ , ½, ½).

To determine how the short range ordering affects the stacking fault energies, we modeled the aluminum distribution near SF plane and calculated $\gamma_{ISF}$ and $\gamma_{US}$ for the different Al positions. First of all, we found that the ordering of two Al atoms in the positions, which correspond to their sites in $D0_3$, results in energy gain of 14 meV as compared to the other possible Al positions. Furthermore, the stacking fault energies $\gamma_{ISF}$ and $\gamma_{US}$ depend on the Al ordering near the SF plane. We obtained that if the positions of two Al atoms at the nearest SF planes correspond to the $D0_3$ ordering, $\gamma_{US}$ is sharply reduced from 230.3 mJ/m$^2$ to 137.3 mJ/m$^2$, while $\gamma_{ISF}$ almost does not change (it decreases by only 4.2 mJ/m$^2$). To understand the reasons for such a selective effect of the Al ordering on the generalized stacking fault energies, we considered the changes in the Fe-Fe, Fe-Al and Al-Al distances under displacements on the (111) plane. The 1/12<112> displacement, which corresponds to unstable stacking fault ($\gamma_{US}$) makes two ordered Al atoms being at the distance $R_{Al-Al}$ of 2.89 Å, which is larger than the Al-Al distance of 2.82 Å in fcc Al, while for all other Al positions $R_{Al-Al}$ becomes equal to 2.33 Å (that is much shorter than in fcc Al).  So, the bonding between Al atoms is much stronger in the second structure, while the structure with partial Al ordering corresponds to the weaker Al-Al bonding with respect to those in fcc Al.

Three Al atoms ordered within three layers near the SF plane lead to a further reduction of $\gamma_{US}$ up to 83 mJ/m$^2$, while $\gamma_{ISF}$ decreases to -356 mJ/m$^2$.  Thus, short range order reduces the energy barrier $\gamma_{US}$ for the stacking fault dislocations, but does not increase the SFE. Besides, short range ordering may be responsible for the overestimated SFE in our calculations for non-ordered Al distribution. Based on these results we can conclude that a short range ordering, which was suggested in Fe–Mn–Al–C steel to explain the planar glide deformation before occurrence of mechanical twinning regardless of the SFE value [10], is related to the precursor of $Fe_3AlC_x$ precipitates. We demonstrate that the mechanism of the softening phenomenon is attributed to a sharp lowering of the energy barrier for stacking fault dislocations due to a short range ordering of aluminum atoms in Fe-Al-C.

## 4. Conclusions

*Ab-initio* calculations were performed to find the relative distribution of C, Mn and Al impurities in fcc Fe matrix. We predict that carbon atoms repel each other and avoid the occupation of the nearest interstitial sites. Carbon demonstrates attractive interaction with manganese and the Mn-C binding energy increases with Mn content. Aluminum prefers to substitute for the second nearest iron atoms with respect to carbon and the predicted distribution of Al and C in austenite exactly corresponds to the ordered structure of *k*-carbide $Fe_3AlC$.

The calculated generalized stacking fault energies demonstrate that only impurities distributed within the two layers near stacking fault plane affect the SFE. We showed that the stacking fault energy does not depend on the average concentration of the impurities in the matrix, and found a strong sensitivity of the SFE on the concentration of impurities in the vicinity of stacking fault defect. The correct parabolic dependence of the SFE on manganese concentration was obtained for manganese at the SF plane, while manganese located more than one interlayer distance away from the SF plane does not influence the staking fault energy. This is one of the possible reasons why the consideration of disordered structure may fail to predict SFE. Our results correlate with the phase diagram of Fe-Mn which demonstrates that ε –martensite is not formed above 30% Mn. The lowering of SFE with Mn content up to 25 at.% also demonstrates that the location of manganese near the stacking faults is energetically favorable compared to being far away from this region. Both aluminum and carbon linearly increase SFE (both impurities suppress the γ → ε transformation) and this effect is more pronounced for the interstitial carbon. We show that carbon prefers to occupy a more distant position from the stacking fault region. Short range order in the Fe-Al-C alloys strongly affects the energy barrier for the dislocation nucleation and may result in softening, while it inhibits the increase in SFE.

**Acknowledgements**

The work is supported by the National Science Foundation (grant CMMI-0726888). N.I.M. also acknowledges the support from the Russian Foundation for Basic Research (Grant No. 09-03-00070a)

Table 1. Preference site energies (meV) for impurities in fcc Fe-Mn-Al-C. Zero energy corresponds to configuration with the lowest energy. Impurity sites are shown in Fig.1,2.

| Site | Fe32C2 | Fe31MnC | Fe30Mn2C | Fe31AlC | Fe30MnAlC |
|---|---|---|---|---|---|
| 1 | 268 | 0 | 0 | 282 | 259 |
| 2 | 1444 | 26 | 21 | 0 | 297 |
| 3 | 65 | 24 | 56 | 245 | 0 |
| 4 | 182 | 29 | 55 | 187 | 225 |
| 5 | 0 | | 72 | | 307 |
| 6-11 | | | 42-85 | | 84-190 |

Table 2. The SF energies (mJ/m$^2$) for single impurity in the n-layer below SF and two impurity atoms in *k*- and *l*- layers ($n = k,l$).

|   |   | $n = 0$ | $n = 1$ | $n = 2$ | $n = 3$ | $n = 0,0$ |
|---|---|---|---|---|---|---|
| Mn | $\gamma_{US}$ | 423 | 430 | 433 | 435 | 409 |
|    | $\gamma_{ISF}$ | -372 | -354 | -349 | -347 | -392 |
| C  | $\gamma_{US}$ | 547 | 541 | 493 | 449 | 601 |
|    | $\gamma_{ISF}$ | 49 | -51 | -145 | -320 | 203 |
| Al | $\gamma_{US}$ | 319 | 432 | 429 | 435 | 232 |
|    | $\gamma_{ISF}$ | -284 | -295 | -343 | -347 | -192 |

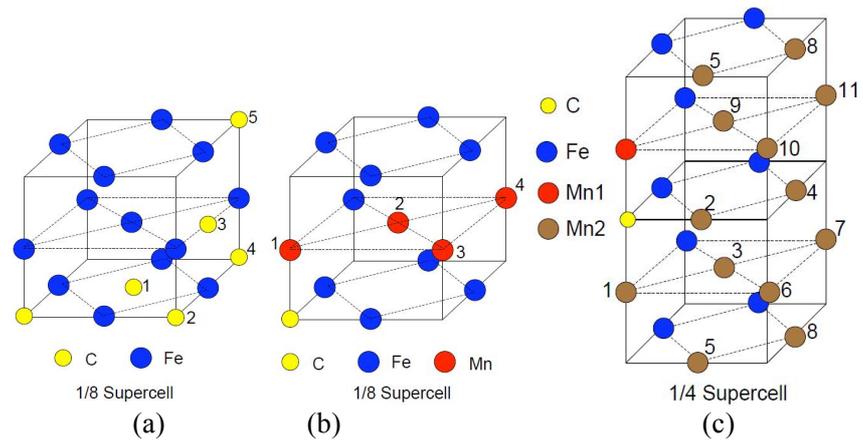

**Figure 1**. (Color online).Occupation sites for (a) carbon atoms and (b,c) carbon and manganese atoms in austenite matrix.

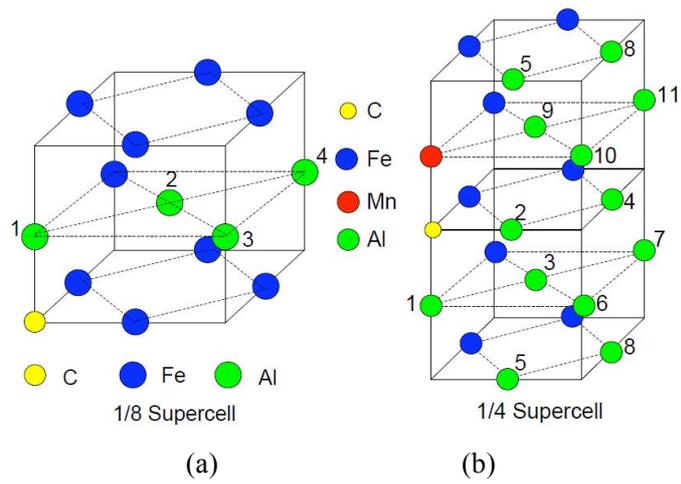

**Figure 2.** (Color online). Occupation sites for aluminum in (a) Fe-Al (sites 1-4) and (b) in Fe-Mn-Al-C (sites 1-11).

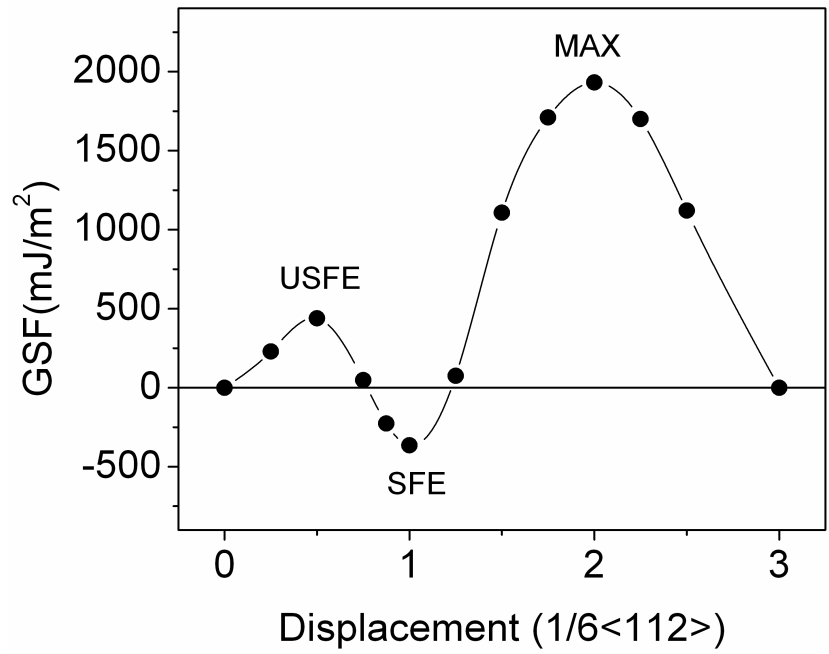

**Figure 3**. The generalized stacking fault (GSF) energies for fcc Fe.

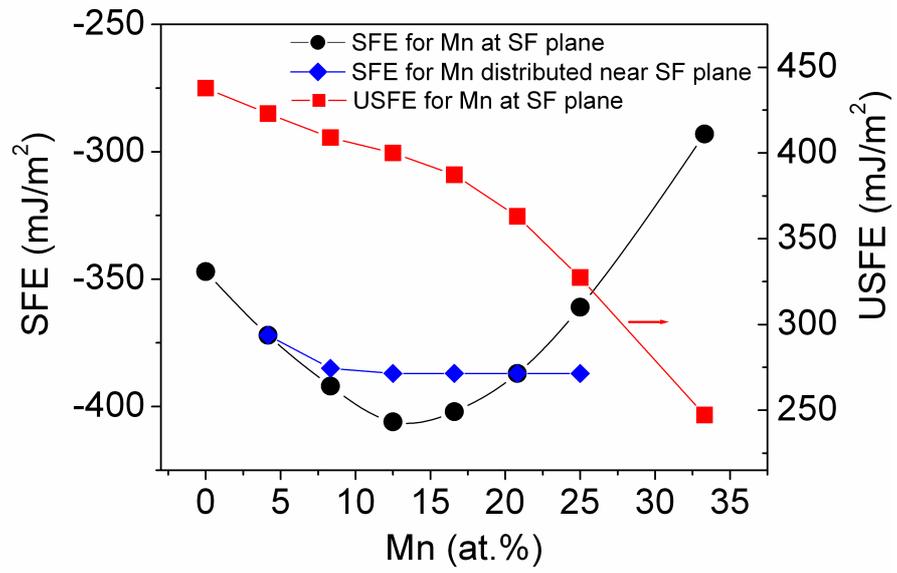

**Figure 4**. (Color online). The intrinsic stacking faut energies (SFE) and unstable stacking fault energies (USFE) as dependent on Mn concentration near stacking fault in fcc Fe-Mn.

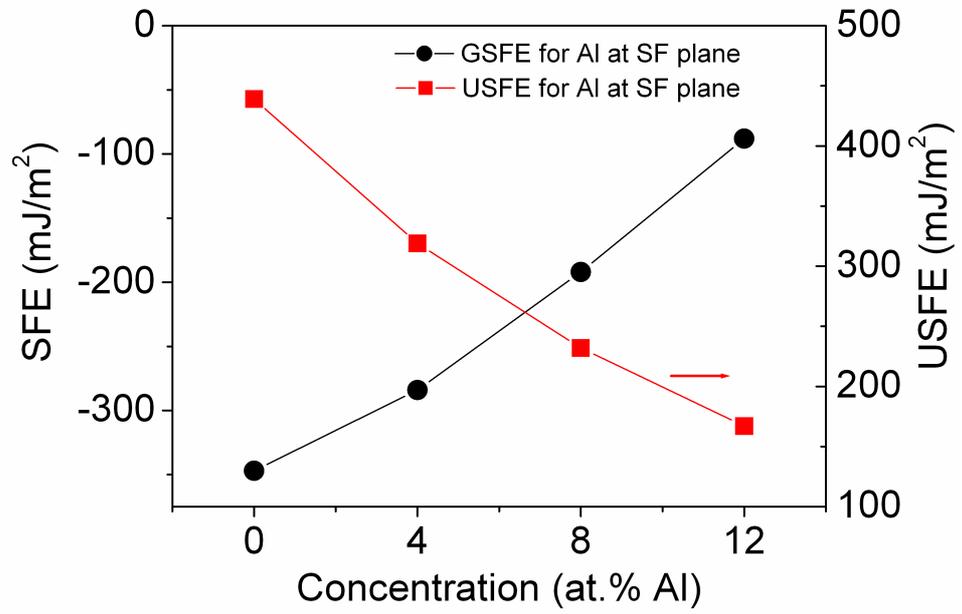

**Figure 5**. (Color online). The intrinsic stacking faut (SFE) and unstable stacking fault (USFE) energies as dependent on Al concentration at stacking fault plane in fcc Fe-Al.